\documentclass[aps,prl,superscriptaddress,twocolumn,amsmath,amssymb]{revtex4}
\usepackage[latin2]{inputenc}
\usepackage{amsmath}
\usepackage{graphicx}
\usepackage{multirow}
\usepackage{dcolumn}
\usepackage{epstopdf}
\usepackage{color}

\newcommand{\ba}{\begin{eqnarray*}}
\newcommand{\ea}{\end{eqnarray*}}
\newcommand{\baa}{\begin{eqnarray}}
\newcommand{\eaa}{\end{eqnarray}}
\def\bar{\begin{array}}
\def\ear{\end{array}}
\def\LB{\left(}
\def\RB{\right)}

\def\u{\uparrow}
\def\d{\downarrow}

\def\s{\sigma}
\def\f{\frac}

\def\nn{\nonumber}
\def\kk{\mathbf{k}}
\def\r{\mathbf{r}}

\begin{document}

\title{Approximate formula for the macroscopic polarization including quantum fluctuations} 

\author{Ryan Requist}
\email{rrequist@mpi-halle.mpg.de}
\affiliation{
Max Planck Institute of Microstructure Physics, Weinberg 2, 06114 Halle, Germany 
}
\author{E. K. U. Gross}
\affiliation{
Max Planck Institute of Microstructure Physics, Weinberg 2, 06114 Halle, Germany
}
\affiliation{
Fritz Haber Center for Molecular Dynamics, Institute of Chemistry, The Hebrew University of Jerusalem, Jerusalem 91904, Israel
}

\date{\today}

\begin{abstract}
The many-body Berry phase formula for the macroscopic polarization is approximated by a sum of natural orbital geometric phases with fractional occupation numbers accounting for the dominant correlation effects.  This reduced formula accurately reproduces the exact polarization in the Rice-Mele-Hubbard model across the band insulator-Mott insulator transition.  A similar formula based on a one-body reduced Berry curvature accurately predicts the interaction-induced quenching of Thouless topological charge pumping.
\end{abstract}

\maketitle

Macroscopic polarization is a fundamental property of dielectric materials from which permittivity and piezoelectric tensors and other physical variables can be derived.  A solid can polarize spontaneously, as occurs in ferroelectrics, or in response to an applied electric field, strain and other external perturbation  \cite{rabe2007}. 

A satisfactory theory of bulk macroscopic polarization was formulated only relatively recently \cite{resta1992,king-smith1993,resta1994} after the realization that  changes in the macroscopic polarization, rather than its nominal value, are the physically relevant and experimentally measurable quantities; see Ref.~\onlinecite{resta2007} for a lucid account.  King-Smith and Vanderbilt derived the following formula for the change induced by adiabatically varying an arbitrary Hamiltonian parameter $\lambda$  \cite{king-smith1993}: 
\begin{align}
\Delta \mathbf{P} =  -e \int_0^1 d\lambda  \int_{BZ} \!\! \f{d^3k}{(2\pi)^3} 2\mathrm{Im} \sum_n^{occ}  \langle \nabla_{\kk} u_{n\kk} | \partial_{\lambda} u_{n\kk} \rangle {,}\label{eq:dP:king-smith:vanderbilt}
\end{align}
where $u_{n\kk}$ is the periodic part of the Bloch state $\chi_{n\kk}$.  The $k$ integral is over the Brillouin zone and the sum is over occupied bands.  The integrand contains a mixed $(\kk,\lambda)$ Berry curvature $B_{\kk\lambda} = 2\mathrm{Im} \sum_n\langle \nabla_{\kk} u_{n\kk} | \partial_{\lambda} u_{n\kk} \rangle$ \cite{berry1984}, which also appears in Thouless charge pumping \cite{thouless1983}.

The change in polarization in the direction of a lattice vector $\mathbf{R}_{\alpha}$ can be expressed as a Berry phase \cite{king-smith1993}, e.g.
\begin{align} 
\Delta P_3 = -\f{e}{(2\pi)^3} \sum_n^{occ} \int \!dk_1 dk_2 \!\int_0^{|\mathbf{G}_3|} \! i \langle u_{n\kk} | \partial_{k_3} u_{n\kk} \rangle dk_3\big|^{\lambda=1}_{\lambda=0} {,}
\label{eq:king-smith-vanderbilt:2}
\end{align}
where $\Delta P_3 = \f{1}{2\pi} \mathbf{G}_3 \cdot \Delta \mathbf{P}$ and the $(k_1,k_2)$ integral is taken over the parallelogram spanned by the reciprocal lattice vectors $\mathbf{G}_1$ and $\mathbf{G}_2$.  The geometric phase of a Bloch state on a path traversing the Brillouin zone was introduced by Zak and related to Wyckoff positions \cite{zak1989,michel1992}. 

The King-Smith--Vanderbilt formula is exact for noninteracting electrons and has given good results for ferroelectric perovskites and other materials \cite{resta1993,zhong1994,posternak1997,ghosez1998,saghi-szabo1998,zhang2017}.  However, if the Bloch states are chosen to be the Kohn-Sham orbitals from a density functional theory (DFT)
calculation, as is usually done, the formula is not guaranteed to yield the exact polarization even if the exact exchange-correlation potential is used \cite{gonze1995}.   
The King-Smith--Vanderbilt formula may give incorrect results in strongly correlated materials, and if the single-particle orbitals are chosen to be the Kohn-Sham orbitals, then it is ill-defined for any insulator whose Kohn-Sham system is metallic \cite{godby1989}.  On the other hand, if the Bloch states are obtained from an exact calculation in current-DFT \cite{vignale1987}, then an adaptation of the arguments in Ref.~\onlinecite{shi2007} suggests that Eq.~(\ref{eq:dP:king-smith:vanderbilt}) will give the correct $\Delta \mathbf{P}$.

Ortiz and Martin \cite{ortiz1994} generalized the King-Smith--Vanderbilt formula to 
correlated many-body systems using twisted boundary conditions, a concept that has been used to analyze the insulating state of matter \cite{kohn1964,kohn1967}, the integer quantum Hall effect \cite{laughlin1981,niu1985} and topological charge pumping \cite{thouless1983,niu1984}.  For one-dimensional systems, the Ortiz-Martin formula reads
\begin{align}
\Delta P = -\f{e}{2\pi} \underset{N/L=const.}{\lim_{N,L\rightarrow\infty}}  \int_{0}^{2\pi/L} \! i\langle \Phi_0 | \partial_k \Phi_0 \rangle dk\big|_{\lambda=0}^{\lambda=1} {,}
\label{eq:dP:ortiz:martin}
\end{align}
where $N$ is the number of electrons in a supercell of length $L$.  The many-body state $\Phi_0=\Phi_0(x_1,\ldots,x_N)$ is the ground state of the ``twisted'' Hamiltonian 
\begin{align}
\hat{H}(k,\lambda) = \sum_{i=1}^N \frac{(p_i+\hbar k)^2}{2m} + \sum_{\langle ij\rangle} \f{e^2}{|r_i-r_j|}+ \hat{V}_{\rm ext}(\lambda) {,}\label{eq:H:twisted}
\end{align}
where $\hat{V}_{\rm ext}(\lambda)$ includes the electron-ion interaction and any other external potentials and $k$ generates an effective magnetic flux that takes on the role of the twisted boundary conditions.  $\Phi_0$ is related to the ground state $\Phi_0^{\prime}$ of the original Hamiltonian by $\Phi_0 = e^{i k(x_1 + x_2 + \cdots + x_N)} \Phi_0^{\prime}$.

The main result we report here is a geometric phase formula for the macroscopic polarization that maintains the simplicity and utility of the King-Smith--Vanderbilt formula while capturing the most important correlations in the Ortiz-Martin result. The reduced formula is
\begin{align}
\Delta \mathbf{P}_{red} &= -e \!\int_0^1 \!d\lambda \!\!\int_{BZ}\!\! \f{d^3k}{(2\pi)^3} 2\mathrm{Im} \sum_{n=1}^{\infty} \langle \nabla_{\kk} v_{n\kk} | \partial_{\lambda} v_{n\kk} \rangle  {,} \label{eq:dP:ours}
\end{align}
where $v_{n\kk}(\r)=\sqrt{f_{n\kk}} e^{i\zeta_{n\kk}} \phi_{n\kk}(\r)$ is the periodic part of the natural orbital Bloch state $\psi_{n\kk}(\r) = e^{i\kk\cdot\r} v_{n\kk}(\r)$; $v_{n\kk}(\mathbf{r})$ is analogous to $u_{n\kk}(\mathbf{r})$ in Eq.~(\ref{eq:dP:king-smith:vanderbilt}) \cite{endnote}.
The natural orbitals $\phi_{n\kk}(\r)$ and occupation numbers $f_{n\kk}$ are eigenfunctions and eigenvalues of the one-body reduced density matrix (rdm) $\rho_1(\mathbf{r}\s,\mathbf{r}^{\prime}\s^{\prime})$ and $\zeta_{n\kk}$ is a phase variable defined below.

Equation~(\ref{eq:dP:ours}) expresses the change in polarization as a sum of single-particle band contributions, like the King-Smith--Vanderbilt formula, but uses natural orbitals instead of Kohn-Sham orbitals.  The natural orbitals are intrinsic variables of the many-body wave function rather than eigenstates of an effective mean-field Hamiltonian.  Since the natural orbital state $\psi_{n\kk}$ contains the factor $f_{n\kk}$ and $0\leq f_{n\kk} \leq 1$ as a result of quantum and thermal fluctuations, each valence band contribution is diminished with respect to the noninteracting case and there are nonvanishing conduction band contributions.  Equation~(\ref{eq:dP:ours}) rests on the assumption that the sum of the natural orbital geometric phases is a good approximation to the geometric phase of the full correlated many-body state.  
Reasons for the accuracy of this approximation will be discussed below after numerical results are reported.

\textit{Polarization in the Rice-Mele-Hubbard model}. Resta and Sorella \cite{resta1995} applied the Ortiz-Martin formula to the Rice-Mele-Hubbard model, also known as the ionic Hubbard model \cite{egami1993,ishihara1994a}, which is a model obtained by adding Hubbard interactions to the Su-Schrieffer-Heeger \cite{su1979} or Rice-Mele \cite{rice1982} models.  It exhibits a quantum phase transition between band insulating and Mott insulating phases at a critical value of the Hubbard parameter $U_c$ with the many-body geometric phase providing an order parameter for the transition \cite{resta1995}.  Subsequent works have used geometric phases to further characterize quantum phase transitions \cite{ortiz1996,gidopoulos2000,carollo2005,zhu2006,cui2008}.  Recently, higher-order cumulants and the total distribution of the polarization have been calculated for the Rice-Mele model \cite{yahyavi2017}.

\begin{figure}[tb!]
\includegraphics[width=0.70\columnwidth]{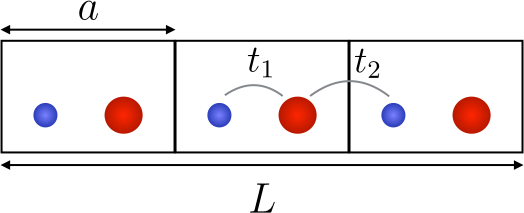}
\caption{Schematic of the Rice-Mele-Hubbard model with one cation (blue) and one anion (red) in the primitive cell.} 
\label{fig:supercell}
\end{figure}

The Hamiltonian of the Rice-Mele-Hubbard model can be written as (see e.g.~Ref.~\onlinecite{asboth2016})
\begin{align}
\hat{H} &= - t_1 \sum_{i\s} c_{i\s}^{\dag} c_{i\s} \otimes \hat{\tau}_x - t_2 \sum_{i\s} \big( c_{i\s}^{\dag} c_{i+1\s} \otimes \f{\hat{\tau}_-}{2} + H.c. \big) \nn \\ 
&\quad+ \Delta \sum_{i\s} c_{i\s}^{\dag} c_{i\s} \otimes \hat{\tau}_z + U \sum_i \hat{n}_{i\u} \hat{n}_{i\d}\otimes \hat{I} {,} \label{eq:rice}
\end{align}
where $\hat{\vec{\tau}}$ are the Pauli matrices in the $(A,B)$ sublattice basis, $\hat{\tau}_{\pm} = \hat{\tau}_x \pm i \hat{\tau}_y$, and $t_{1,2} = t_0 \mp 2g \xi$ with $g=10$eV$/a$ denoting the electron-phonon coupling and $\xi$ the displacement of $B$ with respect to $A$.  As illustrated in Fig.~\ref{fig:supercell}, there are two atoms in the primitive cell; $A$ represents a cation and $B$ an anion.  The lattice constant is $a$.

Before testing Eq.~(\ref{eq:dP:ours}), we calculate the exact $\Delta P$ using the Ortiz-Martin formula, as done by Resta and Sorella.  To apply the Ortiz-Martin formula, one proceeds as follows.  First, choose a supercell of length $L=Ma$, $M\in\mathbb{Z}$, and define the lattice analog of the twisted Hamiltonian in Eq.~(\ref{eq:H:twisted}) by making the replacement 
\begin{align*}
t_2 \big( c_{M\s}^{\dag} c_{1\s} \otimes \f{\hat{\tau}_-}{2}+ H.c. \big) \rightarrow t_2 \big( e^{i\alpha} c_{M\s}^{\dag} c_{1\s} \otimes \f{\hat{\tau}_-}{2}+ H.c. \big)
\end{align*}
in Eq.~(\ref{eq:rice}).  Second, calculate the ground state $|\Phi_0(\alpha)\rangle$ for all values of the twist angle $\alpha=kL \in[0,2\pi]$ using the Lanczos algorithm.  Third, evaluate the geometric phases 
\begin{align}
\gamma(\xi) = \int_0^{2\pi} i \langle \Phi_0 | \partial_{\alpha} \Phi_0 \rangle d\alpha {.}
\label{eq:gamma:exact} 
\end{align}
Here, the discretized Berry phase formula can be used with the boundary condition $\Phi_0(2\pi) = e^{-i2\pi \hat{X}/L} \Phi_0(0)$. 
Fourth, calculate $\Delta P = -(e/2\pi) [\gamma(\xi_2) - \gamma(\xi_1)]$ for an adiabatic variation from $\xi_1$ to $\xi_2$ for a series of $L$ values and extrapolate to the thermodynamic limit. 

Returning now to Eq.~(\ref{eq:dP:ours}), we observe that it cannot be tested straightforwardly because we do not have a way to calculate the exact $v_{nk}$ of the infinite Rice-Mele-Hubbard model.  However, from the exact ground state $|\Phi_0(\alpha)\rangle$ of a supercell of length $L$, obtained as described above, it is straightforward to calculate the natural orbitals $\phi_n$, occupation numbers $f_n$ and phases $\zeta_n$ as functions of $\alpha$.
Hence, we define the $L$-dependent {\it reduced} polarization
\begin{align}
\Delta P_{red}(L) = -\f{e}{2\pi} \sum_n &\bigg[ \int_0^{2\pi m} i \langle v_n | \partial_{\alpha} v_n \rangle d\alpha \bigg]_{\xi_1}^{\xi_2} {,} \label{eq:dP:model}
\end{align}
which converges to Eq.~(\ref{eq:dP:ours}) in the thermodynamic limit.  Since the convergence is rapid (based on extrapolation, the difference between $L=5a$ and $L=6a$ is $\lesssim 2\%$),
we will simply compare the results of Eqs.~(\ref{eq:dP:ortiz:martin}) and (\ref{eq:dP:model}) for finite $L$.  The sum over $n$ in Eq.~(\ref{eq:dP:model}) runs over the unfolded natural orbital bands and $m$ is the smallest integer needed to unfold them (see below).

Defining the natural orbital geometric phases
\begin{align}
\gamma_{n}(\xi) = \! \int_0^{2\pi m} \!i \langle v_n| \partial_{\alpha} v_n \rangle d\alpha 
{,}
\label{eq:gamma:reduced}
\end{align}
it is seen that Eq.~(\ref{eq:dP:model}) depends on the sum $\sum_n\gamma_{n}$, which we shall refer to as the (one-body) reduced geometric phase 
$\gamma_{red}$, since it approximates the many-body geometric phase in Eq.~(\ref{eq:gamma:exact}) with the variables from reduced density matrices. 

Since the one-body part of the Hamiltonian, $\hat{H}_0$, has translational symmetry, i.e.~$[\hat{H}_0, \hat{T}_a]=0$, where $\hat{T}_a$ is the displacement operator, its eigenstates are readily obtained by diagonalizing the $k$-space Hamiltonian
\begin{align}
\hat{H}_0(k) = \langle k\s | \hat{H}_0 | k\s \rangle = \vec{h}_0(k\s) \cdot \hat{\vec{\tau}} {,}
\label{eq:h:kspace}
\end{align}
where $\vec{h}_0(k\s) = \{ -t_1 - t_2 \cos k , -t_2 \sin k, \Delta \}$.  Here, the plane waves $|k\s\rangle = \f{1}{\sqrt{M}} \sum_{l=1}^M e^{ikl} c_{l\s}^{\dag}|0\rangle$ are defined for periodic boundary conditions and $k=2\pi n/M$; $n=0,1,\ldots M-1$.  The eigenfunctions of $\hat{H}_0(k)$ define the periodic parts $|u_{nk}\rangle$ of the Bloch states $|\chi_{nk\s}\rangle = |u_{nk}\rangle |k\s\rangle$.  

When the artificial gauge potential implied by $\alpha$ is turned on, the states maintain their Bloch form but the allowed values of $k$ shift to $k=(2\pi n+\alpha)/N$.  Hubbard interactions do not break overall translational symmetry, so the many-body eigenstates can be labeled by the total quasimomentum $K$.  Since the Hamiltonian commutes with $\hat{S}^2$ and $\hat{S}_z$, we also have the quantum numbers $S$ and $S_z$.  For example, only configurations whose occupied Bloch states $\{|\chi_{n_ik_i\s_i}\rangle\}$ satisfy $\sum_{i=1}^N k_i=K$ and $\sum_{i=1}^N \s_i=S_z$ contribute to the many-body eigenstate.  

The results presented in the following were obtained for the Rice-Mele-Hubbard model with $N=6$ and $L=3a$, corresponding to 6 electrons in 6 sites.  The ground state is a spin singlet with quantum numbers $K=0$, $S=0$ and $S_z=0$.  The dimension of the $S_z=0$ Hilbert space is 400, which reduces to 136 with $K=0$ \cite{suppl}.  The SNEG package was used to set up the Hamiltonian \cite{zitko2011}.

After calculating the ground state $|\Phi_0\rangle$, the natural orbitals and occupation numbers were readily obtained by diagonalizing the one-body rdm
\begin{align}
\rho_{aa'^{\prime}}(k\s) = \langle \Phi_0 | \hat{c}_{a^{\prime} k\s}^{\dag} \hat{c}_{ak\s} | \Phi_0 \rangle {,}  
\end{align}
where $\hat{c}_{a k\s}^{\dag}$ is the creation operator for the sublattice Bloch state $\chi_{ak\s} = |a\rangle|k\s\rangle$, where $a=A,B$.  Since the one-body rdm commutes with $\hat{T}_a$ and $\hat{S}_z$, it is diagonal in $k$ and $\s$.  

\begin{figure}[tb!]
\includegraphics[width=0.9\columnwidth]{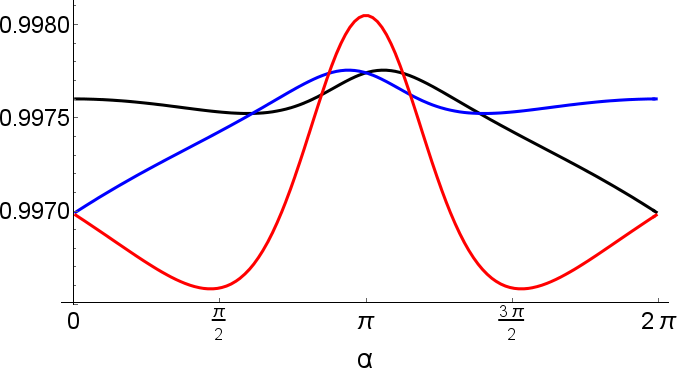}
\caption{Largest three occupation numbers $f_{n\s}$ for parameters $t_0=3.5$eV, $\xi=0.0140a$, $\Delta=2.0$~eV and $U=0.8t_0$.}
\label{fig:occnums}
\end{figure}

Figure \ref{fig:occnums} shows the occupation number band structure.  The three largest spin-independent occupation numbers $f_1$, $f_2$ and $f_3$ are plotted as functions of $\alpha$; $f_n\equiv f_{n\s}$.  There are additionally three weakly occupied occupation numbers, $f_4$, $f_5$ and $f_6$, which, however, are not independent due to the conditions $f_1+f_6=1$, $f_2+f_5=1$ and $f_3+f_4=1$ \cite{suppl}.  Since the occupation numbers tend to cluster near 0 and 1, there are inevitably frequent crossings as $\alpha$ is varied.  To identify the individual bands, we used the overlap of natural orbitals at adjacent $\alpha$ points.  

\begin{figure}[tb!]
\includegraphics[width=0.93\columnwidth]{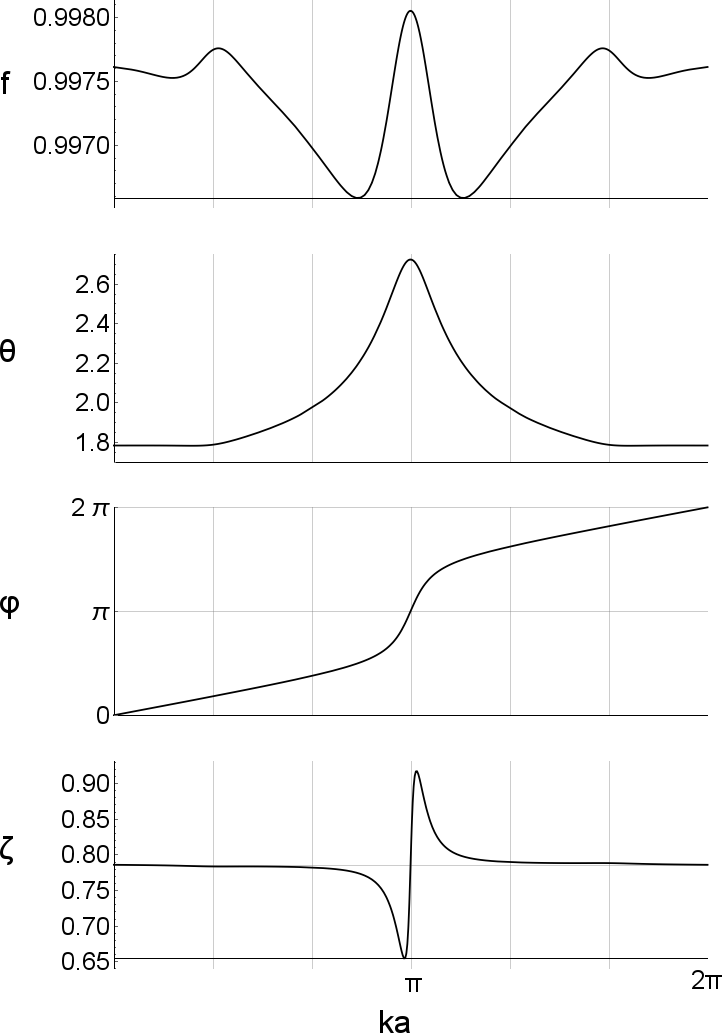}
\caption{Unfolded natural orbital variables $f$, $\theta$, $\varphi$ and $\zeta$ in the normal Brillouin zone for the same parameters as Fig.~\ref{fig:occnums}. }
\label{fig:natorbs}
\end{figure}

The occupation numbers $f_1$, $f_2$ and $f_3$ are branches of a single multivalued function.  The $f_n$ can be matched smoothly to one another at the boundaries of the $\alpha$ domain $[0,2\pi]$, e.g.~$f_1(2\pi)=f_3(0)$, $f_3(2\pi)=f_2(0)$ and $f_2(2\pi)=f_1(0)$.  By extending the domain to $[0,6\pi]$, the occupation numbers $f_1$, $f_2$ and $f_3$ can be ``unfolded'' to form a single strongly occupied valence band in the normal Brillouin zone, as shown in Fig.~\ref{fig:natorbs}.  Similarly, $f_4$, $f_5$ and $f_6$ form a single weakly occupied conduction band.  

In the sublattice basis, the natural orbitals are 
\begin{align}
\phi_n(l) = \left\{ \bar{ll} \f{e^{ik_n l} }{\sqrt{3}} \LB \bar{l} 
\cos(\theta_n/2) \\
\sin(\theta_n/2) e^{i\varphi_n} 
\ear \RB &  n = 1,2,3  \\[0.3cm]
\f{e^{ik_n l}}{\sqrt{3}} \LB \bar{l} 
\sin(\theta_n/2) \\
-\cos(\theta_n/2) e^{i\varphi_n} 
\ear \RB &  n = 4,5,6 \ear \right.
\label{eq:NO}
\end{align}
where $l=0,1,2$ labels the cell within the supercell and $k_1 = k_6 =\f{\alpha}{3}$, $k_2 = k_5 =-\f{2\pi}{3} + \f{\alpha}{3}$ and $k_3 = k_4 =\f{2\pi}{3} + \f{\alpha}{3}$.  The natural orbitals match up smoothly at the boundaries of the interval $[0,2\pi]$ in direct correspondence with the occupation numbers.  Unfolding the natural orbitals defines the functions $\theta(\alpha)$ and $\varphi(\alpha)$ shown in Fig.~\ref{fig:natorbs}.

The last quantities we need are the $\zeta_n$.  These phases could be determined from the variational principle, but since this is impractical in problems with large Hilbert spaces, we propose the following alternative route to calculate them.  First, calculate the set of two-body rdm elements $\rho_{ijkl}^{\rho\s\tau\upsilon}=\langle \Phi_0 | \hat{b}_{l\upsilon}^{\dag} \hat{b}_{k\tau}^{\dag} \hat{b}_{i\rho} \hat{b}_{j\s} | \Phi_0 \rangle$, 
where $\hat{b}_{j\s}^{\dag}$ is the creation operator for natural orbital $\phi_{j\s}$.   Then, use the Moore-Penrose pseudoinverse to solve the overdetermined equations 
$\zeta_{i\rho} + \zeta_{j\s} - \zeta_{k\tau} - \zeta_{l\upsilon} = \mathrm{Arg} \rho_{ijkl}^{\rho\s\tau\upsilon}$. 
For $L=3a$, the only nonzero elements of the type $\rho_{iijj}^{\u\d\u\d}$ are $\rho_{1166}^{\u\d\u\d}$, $\rho_{2255}^{\u\d\u\d}$ and $\rho_{3344}^{\u\d\u\d}$ and their Hermitian conjugates.  These elements are sufficient to determine $\zeta_1-\zeta_6$, $\zeta_2-\zeta_5$ and $\zeta_3-\zeta_4$.  The reduced geometric phase $\gamma_{red}$ only depends on these combinations of $\zeta_n$ variables.  The unfolded $\zeta(\alpha)$ is shown in Fig.~\ref{fig:natorbs}.  

The $\zeta_n$ control relative phases between the configurations of the many-body wavefunction.  Changing the $\zeta_n$ changes the $n$-body correlation functions, e.g., the probability of double occupancy $D_a=\langle \Phi_0 | \hat{n}_{a\u}  \hat{n}_{a\d} | \Phi_0\rangle$ on sublattice $a=A,B$, and therefore affects the energy \cite{requist2011}.  

In terms of the unfolded functions $f(k)$, $\theta(k)$, $\varphi(k)$ and $\zeta(k)$ with $k=\alpha/L$, the periodic part $|v_{nk}\rangle$ of the natural orbital Bloch state $|\psi_{nk}\rangle=|v_{nk}\rangle |k\s\rangle$ can be parametrized as
\begin{align}
|v_{1k}\rangle &= \sqrt{f} e^{i\zeta} \LB \bar{l} 
\cos(\theta/2) \\
\sin(\theta/2) e^{i\varphi+ikr_B}
\ear \RB \nn \\
|v_{2k}\rangle & = \sqrt{1-f} e^{-i\zeta} \LB \bar{l} \sin(\theta/2) \\
-\cos(\theta/2) e^{i\varphi+ikr_B}
\ear \RB {,}
\label{eq:v}
\end{align}
for the valence and conduction bands, respectively. Here $r_B=a/2$ is the coordinate of the ion at site $B$ ($r_A\equiv0$).

\begin{figure}[tb!]
\includegraphics[width=0.97\columnwidth]{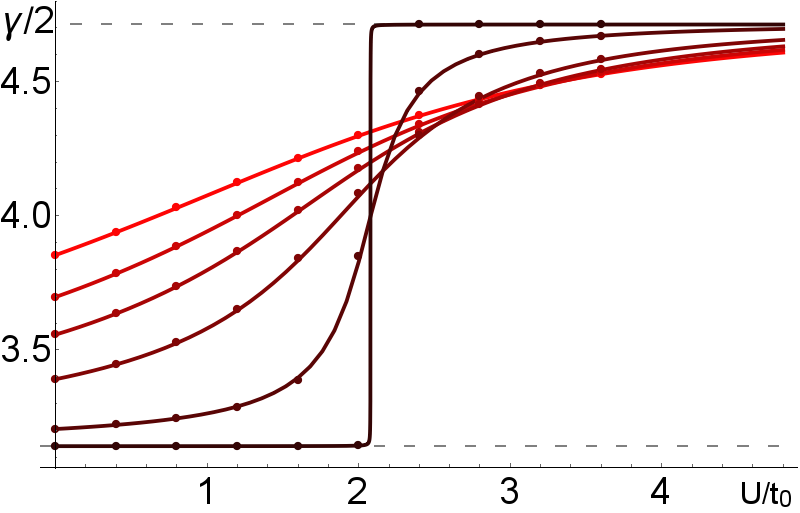}
\caption{Exact (lines) and reduced (points) geometric phases for $\xi=(5.00\times 10^{-6},0.0035,0.0140,0.0245,0.0350,0.0500)a$ [dark to light]. Dashed lines show $\pi$ and $3\pi/2$.}
\label{fig:gamma}
\end{figure}

We now have the ingredients needed to calculate $\Delta P_{red}$ in Eq.~(\ref{eq:dP:model}).  As all the information is contained in the geometric phases, we shall simply compare the reduced geometric phase $\gamma_{red}$ with the exact geometric phase $\gamma$.  In the noninteracting case, $\gamma/2$ is related to the valence band Wannier function center according to $\langle \hat{r} \rangle/a = (\gamma/2)/(2\pi)$; the factor of 1/2 occurs due to the double occupancy of the spin-degenerate band.  Figure \ref{fig:gamma} shows $\gamma/2$ and $\gamma_{red}/2$ as functions of $U$ for several $\xi$.  For $\xi=5\times 10^{-6}a$, corresponding to an almost centrosymmetric lattice, there is an almost discontinuous jump of $\pi$ in $\gamma$ as $U$ passes through $U_c$ \cite{resta1995}.  This implies a sudden change of $e/2$ in the polarization at the band insulator-Mott insulator transition.
The reduced geometric phase is an accurate approximation to the exact geometric phase throughout the range of parameters considered in Fig.~\ref{fig:gamma}.  The calculations were performed with between 64 and 96 $\alpha$ points, and the error $|\gamma_{red}-\gamma|$ \cite{suppl} is on the order of 1\%.  The data points for $\xi=5\times 10^{-6}a$, where obtained for $|v_{nk}\rangle$ without the $\sqrt{f_n}e^{i\zeta_n}$ factors.

The foregoing calculation of the natural orbital states $\psi_{n\kk}$ from the exact many-body state $|\Phi_0\rangle$
is an expedient to avoid confounding errors which would arise in approximate methods that circumvent the calculation of $|\Phi_0\rangle$. The natural orbitals $\phi_{n\kk}$ and occupation numbers $f_{n\kk}$ can be efficiently calculated with reduced density matrix functional theory \cite{gilbert1975,sharma2013}, but that theory does not determine the $\zeta_{n\kk}$.  A generalized functional theory, which would provide the $\psi_{n\kk}$ if adapted to periodic systems, has been introduced \cite{giesbertz2010}.
 
The accuracy of Eq.~(\ref{eq:dP:ours}) is partially a consequence of $N$-representability constraints \cite{coleman1963,klyachko2006,altunbulak2008}, which are nontrivial (in)equalities that the occupation numbers must satisfy in order to be consistent with an $N$-electron pure state.  In some two- and three-electron systems, the exact saturation of these constraints is known to make the many-body geometric phase reduce exactly to the sum of natural orbital geometric phases.  This occurs in the two-site Hubbard model \cite{requist2011} and three-site Hubbard ring \cite{requist2014a} as a consequence of the L\"owdin-Shull \cite{loewdin1956} and Borland-Dennis conditions \cite{borland1972}.  To our knowledge, the $N$-representability constraints are not yet known for the case of interest here, 
i.e.~$N=6$ and Hilbert space dimension $d=12$, although a general algorithm for determining them has been introduced \cite{klyachko2006,altunbulak2008}.  If the inequality constraints are found to be nearly saturated, i.e.~if the occupation numbers are quasipinned \cite{klyachko2009,schilling2013,benavides-riveros2013}, it would suggest that the reduced geometric phase deviates from the full geometric phase by a quantity that vanishes as the occupation numbers approach the relevant boundary of their allowed region.  

Natural orbital geometric phases are themselves bona fide geometric phases that are equally valid for pure and mixed states \cite{requist2012} and hence also apply to systems at finite temperature.  The natural orbital Bloch states $\psi_{n\kk}(\mathbf{r})=e^{i\kk\cdot\mathbf{r}} v_{n\kk}(\mathbf{r})$ can be used to define \textit{natural Wannier functions} 
\begin{align}
|w_{n\mathbf{R}}\rangle =  \int_{BZ} \f{d^3k}{(2\pi)^3} e^{-i\kk\cdot\mathbf{R}} |\psi_{n\kk}\rangle {.}
\label{eq:wannier}
\end{align}
The Wannier function centers $\langle w_{n\mathbf{0}} | \mathbf{r} | w_{n\mathbf{0}} \rangle$ imply band-decomposed contributions to the polarization and Born effective charges, similar to corresponding decompositions for noninteracting electrons \cite{ghosez1998,ghosez2000}.  Unlike conventional Wannier functions \cite{marzari2012}, the $|w_{n\mathbf{R}}\rangle$ are \textit{unique} (up to a trivial relabeling associated with a shift of origin); correlations provide a ``background'' that fixes all $\zeta_{n\kk}$ up to a common $\kk$-independent phase.  Since $\langle w_{n^{\prime} \mathbf{R}^{\prime}} | w_{n\mathbf{R}} \rangle = \int d^3k \,\mathrm{exp}[-i\kk(\mathbf{R}-\mathbf{R}^{\prime})] f_{n\kk} \delta_{nn^{\prime}}/(2\pi)^{3}$, the $|w_{n\mathbf{R}}\rangle$ are not orthonormal and their overlap depends on $f_{n\kk}$.  More details are available in Ref.~\cite{suppl}.

\textit{Thouless charge pumping.} A special case of Eq.~(\ref{eq:dP:ours}) occurs when the adiabatic perturbation parametrized by $\lambda$ is cyclic.  In this case, the pumped charge 
\begin{align}
Q &= \f{e}{2\pi} \int_0^1 d\lambda \int_0^{2\pi/a} dk B_{\lambda k}
\label{eq:Q} {,}
\end{align}
is a topological invariant \cite{thouless1983,niu1984}.  We have calculated $Q$ for the cyclic driving protocol $t_1 = t_0 + (t_0/8) \cos(2\pi\lambda)$, $t_2 = t_0 - (t_0/8) \cos(2\pi\lambda)$ and $\Delta = (t_0/8) \sin(2\pi\lambda)$, 
which pumps charge to the right \cite{vanderbilt1993}. A transition from $Q=2$ to $Q=0$ occurs at $U^*=0.630\pm 0.001t_0$. An approximate calculation using $B_{red,\lambda k}$ instead of $B_{\lambda k}$ in Eq.~(\ref{eq:Q}) gives the transition at $U^*=0.630\pm 0.001t_0$.  Also in the case of \textit{nonadiabatic} charge pumping, there is a contribution that can be approximated in terms of the natural orbital geometric phases \cite{requist2014a}.

The reduced Berry curvature $B_{red,\mu\nu}$ and the symmetry properties of the $\psi_{n\kk}$ states, e.g.~under time-reversal and inversion, are promising quantities for the practical calculation of topological invariants in the presence of interactions and thermal fluctuations, e.g.~in quantum Hall systems \cite{thouless1982,avron1983,kane2005,bernevig2006} and topological insulators \cite{fu2007,moore2007,roy2009,antonius2016,monserrat2016}.  The fact that the $\psi_{n\kk}$ are built from single-particle orbitals suggests they can be efficiently calculated by \textit{ab initio}-based methods.  This points to the possibility of using the $\psi_{n\kk}$ states in realistic calculations of topological Mott insulators and other strongly correlated materials for which DFT runs into difficulty.

\bibliography{bibliography}

\begin{thebibliography}{66}
\bibitem[{end()}]{endnote}
\bibinfo{howpublished}{$|\psi_{n\mathbf{k}}\rangle$ is invariant under phase
  transformations of the $|\phi_{n\mathbf{k}}\rangle$ because
  $\zeta_{n\mathbf{k}} \rightarrow \zeta_{n\mathbf{k}} - \lambda_{n\mathbf{k}}$
  when $|\phi_{n\mathbf{k}}\rangle\rightarrow
  e^{i\lambda_{n\mathbf{k}}}|\phi_{n\mathbf{k}}\rangle$.}
\end{thebibliography}

\end{document}


{\Large \textbf{Supplemental material}}

\section{Many-body wave function}

Here, we characterize the many-body wave function for the case $L=3a$ considered in the main text.  Since $N_{\u}=N_{\d}=3$, the general wave function is 
\begin{align}
|\Psi\rangle = \sum_I c_I |i_1 i_2 i_3\rangle |i_4 i_5 i_6\rangle {,}
\label{eq:psi}
\end{align}
where $I=(i_1,i_2,i_3,i_4,i_5,i_6)$ 
is a multi-index and the first factor is spin-up and the second factor is spin-down,
i.e.~$|i_1 i_2 i_3\rangle |i_4 i_5 i_6\rangle= c_{i_1\u}^{\dag} c_{i_2\u}^{\dag} c_{i_3\u}^{\dag} c_{i_4\d}^{\dag} c_{i_5\d}^{\dag} c_{i_6\d}^{\dag}|0\rangle$.

The exact wave function has the following three types of terms: (i) terms for which both spin up and spin down factors come from the set $\mathcal{A}$; (ii) terms in which the spin up factors come from the set $\mathcal{B}$ and the spin down factors come from the set $\mathcal{C}$; and (iii) terms in which the spin up factors come from the set $\mathcal{C}$ and the spin down factors come from the set $\mathcal{B}$.
The Borland-Dennis-type set $\mathcal{A}$ is
\begin{align}
\mathcal{A} = \{ |123\rangle, |124\rangle, |135\rangle, |145\rangle, |236\rangle, |246\rangle, |356\rangle, |456\rangle \}
\end{align}
and the non-Borland-Dennis-type sets are 
\begin{align}
\mathcal{B} &= \{ |125\rangle, |136\rangle, |146\rangle, |234\rangle, |256\rangle, |345\rangle \} \\[0.2cm]
\mathcal{C} &= \{ |126\rangle, |134\rangle, |156\rangle, |235\rangle, |245\rangle, |346\rangle \} {.}
\end{align}
The states from set $\mathcal{A}$ have quasimomentum $\alpha$, those from $\mathcal{B}$ have quasimomentum $-2\pi/3+\alpha$ and those from $\mathcal{C}$ have quasimomentum $2\pi/3+\alpha$.  The above three types of terms are the only ones with $K=K_{\u}+K_{\d}=2\alpha$.  There are 136 such configurations.  For later use, we enumerate the states using an ordinal index derived from the binary number of the string of Fock state occupation numbers.  There are 64 BD-like states:
\begin{align*}
|0\rangle &= |123\rangle |123\rangle \nn \\
|1\rangle &= |124\rangle |123\rangle \nn \\
|17\rangle &= |135\rangle |123\rangle \nn \\
|19\rangle &= |145\rangle |123\rangle \nn \\
|102 \rangle &= |236\rangle |123\rangle \nn \\
|104 \rangle &= |246\rangle |123\rangle \nn \\
|126 \rangle &= |356\rangle |123\rangle \nn \\
|127 \rangle &= |456\rangle |123\rangle 
\end{align*}
\begin{align*}
|4\rangle &= |123\rangle |124 \rangle \nn \\
|7\rangle &= |124\rangle |124 \rangle \nn \\
|23\rangle &= |135\rangle |124 \rangle \nn \\
|31\rangle &= |145\rangle |124 \rangle \nn \\
|108\rangle &= |236\rangle |124 \rangle \nn \\
|116\rangle &= |246\rangle |124 \rangle \nn \\
|130\rangle &= |356\rangle |124 \rangle \nn \\
|137\rangle &= |456\rangle |124 \rangle 
\end{align*}
\begin{align*}
|41\rangle &= |123\rangle |135 \rangle \nn \\
|46\rangle &= |124\rangle |135 \rangle \nn \\
|69\rangle &= |135\rangle |135 \rangle \nn \\
|76\rangle &= |145\rangle |135 \rangle \nn \\
|146\rangle &= |236\rangle |135 \rangle \nn \\
|153\rangle &= |246\rangle |135 \rangle \nn \\
|183\rangle &= |356\rangle |135 \rangle \nn \\
|186\rangle &= |456\rangle |135 \rangle 
\end{align*}
\begin{align*}
|52\rangle &= |123\rangle |145 \rangle \nn \\
|55\rangle &= |124\rangle |145 \rangle \nn \\
|84\rangle &= |135\rangle |145 \rangle \nn \\
|91\rangle &= |145\rangle |145 \rangle \nn \\
|161\rangle &= |236\rangle |145 \rangle \nn \\
|168\rangle &= |246\rangle |145 \rangle \nn \\
|192\rangle &= |356\rangle |145 \rangle \nn \\
|197\rangle &= |456\rangle |145 \rangle 
\end{align*}
\begin{align*}
|202\rangle &= |123\rangle |236 \rangle \nn \\
|207\rangle &= |124\rangle |236 \rangle \nn \\
|231\rangle &= |135\rangle |236 \rangle \nn \\
|238\rangle &= |145\rangle |236 \rangle \nn \\
|308\rangle &= |236\rangle |236 \rangle \nn \\
|315\rangle &= |246\rangle |236 \rangle \nn \\
|344\rangle &= |356\rangle |236 \rangle \nn \\
|347\rangle &= |456\rangle |236 \rangle 
\end{align*}
\begin{align*}
|213\rangle &= |123\rangle |246 \rangle \nn \\
|216\rangle &= |124\rangle |246 \rangle \nn \\
|246\rangle &= |135\rangle |246 \rangle \nn \\
|253\rangle &= |145\rangle |246 \rangle \nn \\
|323\rangle &= |236\rangle |246 \rangle \nn \\
|330\rangle &= |246\rangle |246 \rangle \nn \\
|353\rangle &= |356\rangle |246 \rangle \nn \\
|358\rangle &= |456\rangle |246 \rangle 
\end{align*}
\begin{align*}
|262\rangle &= |123\rangle |356 \rangle \nn \\
|269\rangle &= |124\rangle |356 \rangle \nn \\
|283\rangle &= |135\rangle |356 \rangle \nn \\
|291\rangle &= |145\rangle |356 \rangle \nn \\
|368\rangle &= |236\rangle |356 \rangle \nn \\
|376\rangle &= |246\rangle |356 \rangle \nn \\
|392\rangle &= |356\rangle |356 \rangle \nn \\
|395\rangle &= |456\rangle |356 \rangle 
\end{align*}
\begin{align*}
|272\rangle &= |123\rangle |456 \rangle \nn \\
|273\rangle &= |124\rangle |456 \rangle \nn \\
|295\rangle &= |135\rangle |456 \rangle \nn \\
|297\rangle &= |145\rangle |456 \rangle \nn \\
|380\rangle &= |236\rangle |456 \rangle \nn \\
|382\rangle &= |246\rangle |456 \rangle \nn \\
|398\rangle &= |356\rangle |456 \rangle \nn \\
|399\rangle &= |456\rangle |456 \rangle {,}
\end{align*}
and there are 72 non-BD-like states: 
\begin{align*}
|13\rangle &= |126\rangle |125 \rangle \nn \\
|25\rangle &= |134\rangle |125 \rangle \nn \\
|111\rangle &= |235\rangle |125 \rangle \nn \\
|118\rangle &= |245\rangle |125 \rangle \nn \\
|132\rangle &= |346\rangle |125 \rangle \nn \\
|38\rangle &= |156\rangle |125 \rangle 
\end{align*}
\begin{align*}
|14\rangle &= |125\rangle |126 \rangle \nn \\
|110\rangle &= |234\rangle |126 \rangle \nn \\
|30\rangle &= |136\rangle |126 \rangle \nn \\
|37\rangle &= |146\rangle |126 \rangle \nn \\
|133\rangle &= |345\rangle |126 \rangle \nn \\
|123\rangle &= |256\rangle |126 \rangle 
\end{align*}
\begin{align*}
|205\rangle &= |126\rangle |234 \rangle \nn \\
|224\rangle &= |134\rangle |234 \rangle \nn \\
|301\rangle &= |235\rangle |234 \rangle \nn \\
|309\rangle &= |245\rangle |234 \rangle \nn \\
|337\rangle &= |346\rangle |234 \rangle \nn \\
|235\rangle &= |156\rangle |234 \rangle 
\end{align*}
\begin{align*}
|44\rangle &= |125\rangle |134 \rangle \nn \\
|140\rangle &= |234\rangle |134 \rangle \nn \\
|66\rangle &= |136\rangle |134 \rangle \nn \\
|74\rangle &= |146\rangle |134 \rangle \nn \\
|176\rangle &= |345\rangle |134 \rangle \nn \\
|151\rangle &= |256\rangle |134 \rangle 
\end{align*}
\begin{align*}
|51\rangle &= |126\rangle |136 \rangle \nn \\
|68\rangle &= |134\rangle |136 \rangle \nn \\
|147\rangle &= |235\rangle |136 \rangle \nn \\
|154\rangle &= |245\rangle |136 \rangle \nn \\
|182\rangle &= |346\rangle |136 \rangle \nn \\
|81\rangle &= |156\rangle |136 \rangle 
\end{align*}
\begin{align*}
|208\rangle &= |125\rangle |235 \rangle \nn \\
|303\rangle &= |234\rangle |235 \rangle \nn \\
|230\rangle &= |136\rangle |235 \rangle \nn \\
|237\rangle &= |146\rangle |235 \rangle \nn \\
|339\rangle &= |345\rangle |235 \rangle \nn \\
|316\rangle &= |256\rangle |235 \rangle 
\end{align*}
\begin{align*}
|60\rangle &= |126\rangle |146 \rangle \nn \\
|83\rangle &= |134\rangle |146 \rangle \nn \\
|162\rangle &= |235\rangle |146 \rangle \nn \\
|169\rangle &= |245\rangle |146 \rangle \nn \\
|191\rangle &= |346\rangle |146 \rangle \nn \\
|96\rangle &= |156\rangle |146 \rangle 
\end{align*}
\begin{align*}
|217\rangle &= |125\rangle |245 \rangle \nn \\
|318\rangle &= |234\rangle |245 \rangle \nn \\
|245\rangle &= |136\rangle |245 \rangle \nn \\
|252\rangle &= |146\rangle |245 \rangle \nn \\
|348\rangle &= |345\rangle |245 \rangle \nn \\
|331\rangle &= |256\rangle |245 \rangle 
\end{align*}
\begin{align*}
|266\rangle &= |126\rangle |345 \rangle \nn \\
|276\rangle &= |134\rangle |345 \rangle \nn \\
|362\rangle &= |235\rangle |345 \rangle \nn \\
|369\rangle &= |245\rangle |345 \rangle \nn \\
|385\rangle &= |346\rangle |345 \rangle \nn \\
|289\rangle &= |156\rangle |345 \rangle 
\end{align*}
\begin{align*}
|267\rangle &= |125\rangle |346 \rangle \nn \\
|361\rangle &= |234\rangle |346 \rangle \nn \\
|281\rangle &= |136\rangle |346 \rangle \nn \\
|288\rangle &= |146\rangle |346 \rangle \nn \\
|386\rangle &= |345\rangle |346 \rangle \nn \\
|374\rangle &= |256\rangle |346 \rangle 
\end{align*}
\begin{align*}
|223\rangle &= |126\rangle |256 \rangle \nn \\
|248\rangle &= |134\rangle |256 \rangle \nn \\
|325\rangle &= |235\rangle |256 \rangle \nn \\
|333\rangle &= |245\rangle |256 \rangle \nn \\
|355\rangle &= |346\rangle |256 \rangle \nn \\
|259\rangle &= |156\rangle |256 \rangle 
\end{align*}
\begin{align*}
|62\rangle &= |125\rangle |156 \rangle \nn \\
|164\rangle &= |234\rangle |156 \rangle \nn \\
|90\rangle &= |136\rangle |156 \rangle \nn \\
|98\rangle &= |146\rangle |156 \rangle \nn \\
|194\rangle &= |345\rangle |156 \rangle \nn \\
|175\rangle &= |256\rangle |156 \rangle 
\end{align*}

\section{Reduced geometric phase}

Given definitions of the many-body geometric phase 
\begin{align}
\gamma = \int_0^{2\pi} i \langle \Phi | \partial_{\alpha} \Phi \rangle d\alpha
\end{align}
and the reduced geometric phase
\begin{align}
\gamma_{red} = \sum_i \int_0^{2\pi} i \langle v_i | \partial_{\alpha} v_i \rangle d\alpha
\end{align}
with $|v_i\rangle = e^{i\zeta_i} \sqrt{f_i} |\phi_i\rangle$, we derive a formula for the error $\gamma - \gamma_{red}$.  First, write the many-body wave function as a sum over configurations in the natural orbital basis as
\begin{align}
|\Phi\rangle = \sum_I c_I |i_1 i_2 \ldots i_N\rangle {,}
\end{align}
where $I=(i_1,i_2,\ldots,i_N)$ is a multi-index.
Then
\begin{widetext}
\begin{align*}
\langle \Phi | \partial_{\alpha} \Phi \rangle &= \sum_I c_I^* \partial_{\alpha} c_I + \sum_I \sum_{J\neq I} c_J^* c_I \langle J|\partial_{\alpha}I\rangle  \\
&= \sum_I c_I^* \partial_{\alpha} c_I + \sum_I \sum_{J\neq I} \sum_{i\in I} \sum_{j\in J} c_J^* c_I \langle \phi_j| \partial_{\alpha} \phi_i \rangle \langle J| c_j^{\dag} c_i |I\rangle \\
&= \sum_I c_I^* \partial_{\alpha} c_I +  \sum_{i} \sum_{j} \langle \phi_j| \partial_{\alpha} \phi_i \rangle \underbrace{\sum_I^{(i)} \sum_{J\neq I}^{(j)} c_J^* c_I  \langle J| c_j^{\dag} c_i |I\rangle}_{\rho_{ij}} \\
&= \sum_I c_I^* \partial_{\alpha} c_I +  \sum_{i} f_i \langle \phi_i| \partial_{\alpha} \phi_i \rangle \\
&= i \sum_i f_i \partial_{\alpha} \zeta_i  +  \sum_{i} f_i \langle \phi_i| \partial_{\alpha} \phi_i \rangle + \mathcal{E} \\
&= \sum_{i} \langle v_i| \partial_{\alpha} v_i \rangle + \mathcal{E} {,}
\end{align*}
\end{widetext}
where $\sum_I^{(i)}$ denotes the sum over all multi-indices for which one of the $i_n$ is equal to $i$ and we introduced the definition
\begin{align}
\mathcal{E} = \sum_I c_I^* \partial_{\alpha} c_I - i\sum_i f_i \partial_{\alpha} \zeta_i {.}
\end{align}
The error in the reduced geometric phase is 
\begin{align}
\gamma - \gamma_{red} = \int_0^{2\pi} i \mathcal{E} d\alpha {.}
\end{align}

\section{Unique Wannier functions}

Equation (14) of the main text defines general single-particle Wannier-like functions, which are unique in contrast to conventional Wannier functions.  The uniqueness of these natural Wannier functions follows from the fact that the natural Bloch orbitals $|\psi_{n\mathbf{k}\s}\rangle$ from which they are constructed have unique phases. 

The Rice-Mele-Hubbard model has two natural orbital bands -- a strongly occupied (``valence'') band with occupation number $f_{n\mathbf{k}\s}\approx 1$ and a weakly occupied (``conduction'') band with occupation number $f_{n\mathbf{k}\s}\approx 0$.  We have explicitly constructed the corresponding valence-band and conduction-band natural Wannier functions and verified that they satisfy the orthogonality relation $\langle w_{n^{\prime}\R^{\prime}\s^{\prime}}|w_{n\R\s}\rangle = \delta_{nn^{\prime}} \delta_{\s\s^{\prime}}$, where $n$ is the band index.  We have also verified that the change in polarization induced by an adiabatic variation of the dimerization parameter $\xi$ from $\xi_1$ to $\xi_2$ and calculated from the shifts of the natural Wannier function centers according to the formula
\begin{align}
\Delta P \!=\! -e \sum_{n\s} \!\!\bigg[ \langle w_{n0\s} | \hat{x} | w_{n0\s} \rangle\big|_{\xi_2} \!\!-\! \langle w_{n0\s} | \hat{x} | w_{n0\s} \rangle\big|_{\xi_1} \bigg] 
\end{align}
agrees with $\Delta P_{red}$ calculated by Eq.~(8) of the main text.

Figures~\ref{fig:wannier:valence} and \ref{fig:wannier:conduction} show the valence-band and conduction-band natural Wannier functions for $L=3a$ and 128 $k$-points.  The natural Wannier functions constructed from finite supercell calculations have residual gauge dependence, tracing back to the arbitrary $\alpha$-dependent phase of the many-body state $|\Phi_0(\alpha)\rangle$.  However, this is an artifact of the finite size of the supercell and disappears in the limit $L\rightarrow\infty$.  In this limit, all of the natural orbital bands are downfolded to the $\Gamma$-point and have unique relative phases.  Hence, the $\kk$-dependent phases of the $|\psi_{n\kk\s}\rangle$ states are unique for infinite periodic solids.

The unique natural Wannier functions for the Rice-Mele-Hubbard model with $U/t_0=0.4$ are compared with the conventional Wannier functions for the noninteracting Rice-Mele model in Fig.~\ref{fig:wannier:comparison}.  The main visible effect of interactions is to decrease the density imbalance of the natural Wannier functions between sites $A$ and $B$ with respect to that of the conventional Wannier functions.  For simplicity, the nonunique phase of the conventional Wannier functions was chosen such that $\mathrm{Arg}\langle A|u_{nk}\rangle = 0$.

For sufficiently strong dimerization $\xi$, the natural Wannier functions are exponentially localized, i.e.~they display exponential decay $w_{n0\s}(r) \sim e^{-\lambda r}$ with a decay constant $\lambda$.  The decay constants for valence and conduction bands are plotted in Fig.~\ref{fig:decay} as a function of $\xi$ for $U/t_0=(0.4,0.8,1.6)$.  Increasing the dimerization or increasing the Hubbard interaction generally increases the localization of the natural Wannier functions. \\

For small $\xi$, the decay crosses over from exponential to power law with a power $\sim r^{1/2}$.  This crossover coincides with a change in the winding number of the $|\psi_{n\mathbf{k}\s}\rangle$ bands.

\begin{widetext}

\begin{figure}[h!]
\includegraphics[width=0.45\columnwidth]{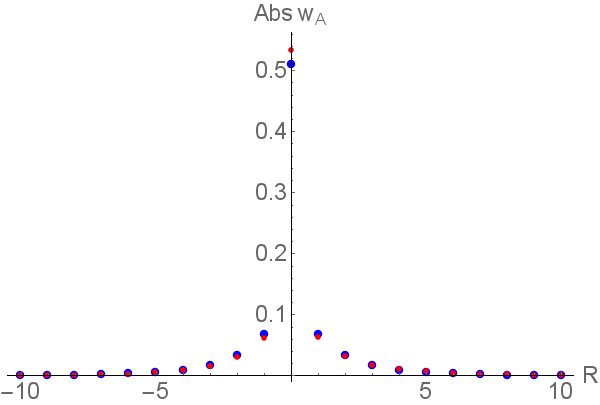} \includegraphics[width=0.45\columnwidth]{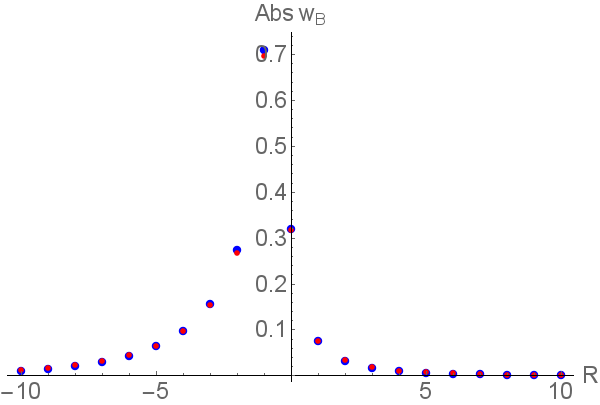}
\caption{Left figure: (red) Modulus of the natural valence-band Wannier function at site $A$ for parameters $t_0=3.5$eV, $\xi=0.0245a$, $\Delta=2$~eV and $U=0.4t_0$, and (blue) modulus of the conventional valence-band Wannier function at site $A$ for the same parameters but $U=0$.  $R$ denotes the cell position.  Right figure: same quantities for site $B$.}
\label{fig:wannier:comparison}
\end{figure}

\begin{figure}[h!]
\includegraphics[width=0.65\columnwidth]{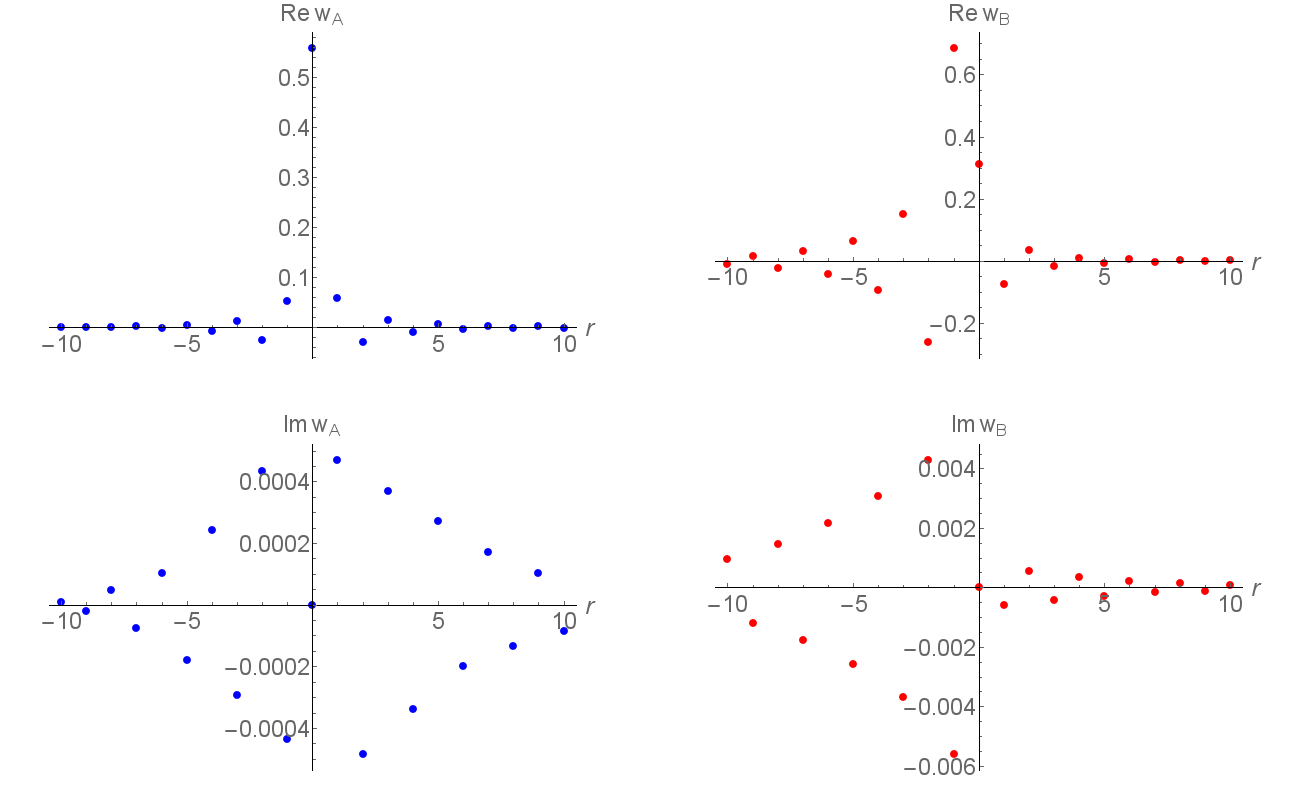}
\caption{Valence-band natural Wannier functions for parameters $t_0=3.5$eV, $\xi=0.0245a$, $\Delta=2$~eV and $U=0.8t_0$. Real and imaginary parts of the amplitude of the Wannier function on sublattice A/B (left/right column).  $r$ denotes the cell position.}
\label{fig:wannier:valence}
\end{figure}

\begin{figure}[h!]
\includegraphics[width=0.65\columnwidth]{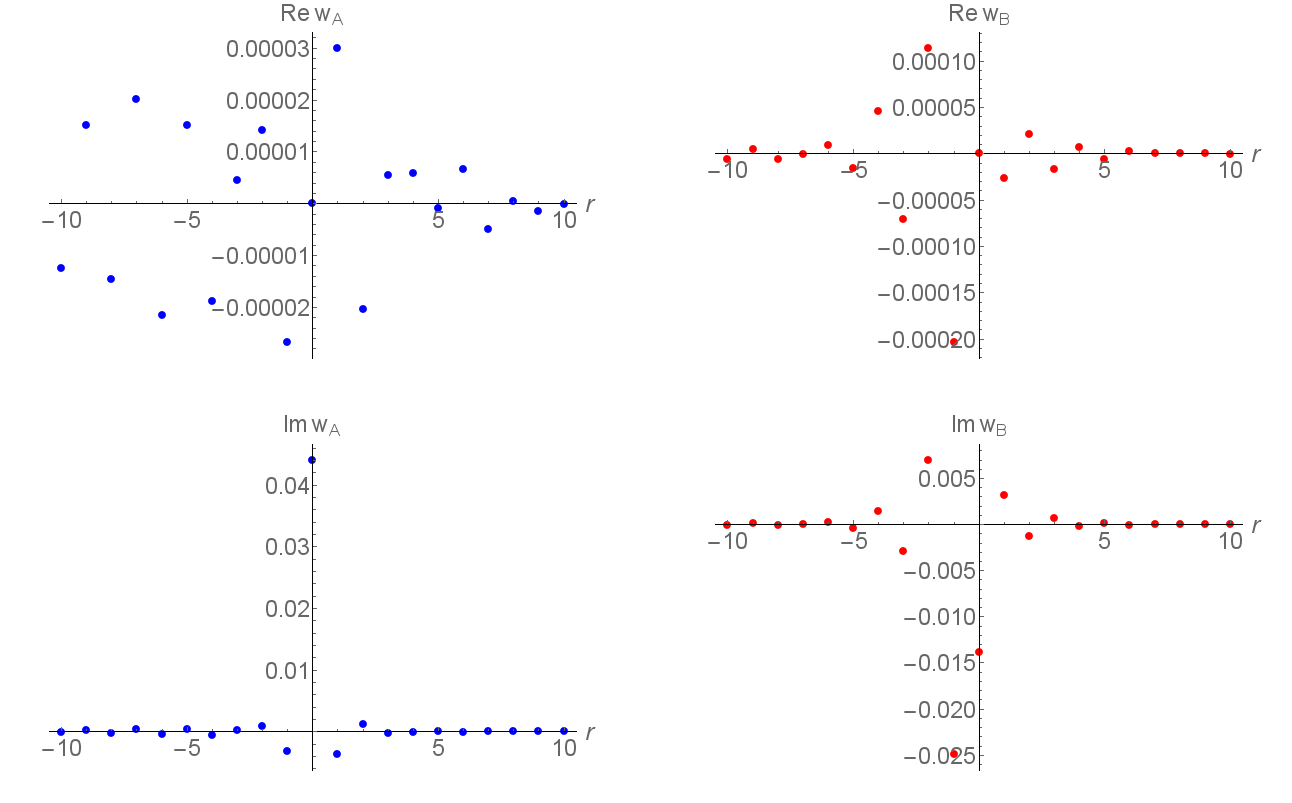}
\caption{Conduction-band natural Wannier functions for parameters $t_0=3.5$eV, $\xi=0.0245a$, $\Delta=2$~eV and $U=0.8t_0$. Real and imaginary parts of the amplitude of the Wannier function on sublattice A/B (left/right column).  $r$ denotes the cell position.}
\label{fig:wannier:conduction}
\end{figure}

\begin{figure}[tb!]
\includegraphics[width=0.45\columnwidth]{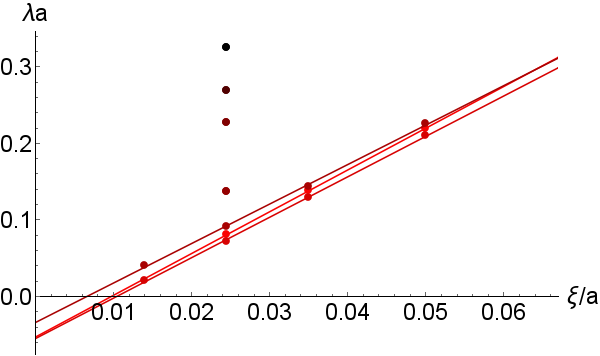}\hspace{0.8cm} \includegraphics[width=0.45\columnwidth]{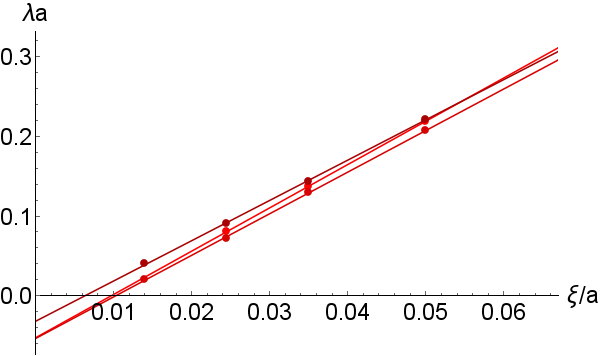}
\caption{Exponential decay constants for the valence-band (left) and conduction-band (right) natural Wannier functions for parameters $t_0=3.5$eV, $\Delta=2$~eV and $U/t_0=(0.4,0.8,1.6)$ (light to dark red) with regression lines.  For the valence band, additional points are plotted for $U/t_0=(2.4,3.2,4.0,4.8)$ (dark red to black).}
\label{fig:decay}
\end{figure}
\end{widetext}

\section{Natural orbital symmetries}

Here we mention the consequences of some symmetries beyond $\hat{\mathbf{S}}^2$, $\hat{S}_z$ and total quasimomentum $\hat{K}$ of the Hamiltonian in Eq.~(6).  

Although the twisted Hamiltonian $\hat{H}(\alpha)$ is not time-reversal invariant for nonzero $\alpha$, time-reversal symmetry relates the ground state $|\Phi_0(\alpha)\rangle$ of $\hat{H}(\alpha)$ to the ground state $|\tilde{\Phi}_0(\alpha)\rangle$ of $\hat{\tilde{H}}(\alpha) = \hat{\mathcal{T}} \hat{H}(\alpha) \hat{\mathcal{T}}^{-1} = \hat{H}(-\alpha)$ according to $|\tilde{\Phi}_0(\alpha)\rangle = \hat{\mathcal{T}}|\Phi_0(\alpha)\rangle$, where $\hat{\mathcal{T}}$ is the time reversal operator, whose action on a single-particle state can be represented as $\hat{\mathcal{T}}=-i\hat{\sigma}_2 \hat{C}$ with $\hat{C}$ being the operator for complex conjugation.  Hence, the natural orbitals, occupation numbers and $\zeta_n$ phases satisfy a similar relationship. For the unfolded $|\psi_{n,\mathbf{k},\s}\rangle$ states, this can be expressed as
\begin{align}
\hat{\mathcal{T}}|\psi_{n,\mathbf{k},\s}\rangle = \mathrm{sgn}(\overline{\s})|\psi_{n,-\mathbf{k},\overline{\s}}^*\rangle {.}
\end{align}

Additionally, the model has a particle-hole symmetry responsible for the symmetry of the occupation number bands with respect to a vertical reflection about $f_{n,\mathbf{k},\s}=\f{1}{2}$.  In other words, if we define the eigenvalues $g_{n,\mathbf{k},\s} = f_{n,\mathbf{k},\s}-\f{1}{2}$ of the reduced density matrix $\Delta\rho(\mathbf{k},\s) = \rho(\mathbf{k},\s)-\f{1}{2}I$, then for every band with $g_{n,\mathbf{k},\s}$ there is a corresponding band with $-g_{n,\mathbf{k},\s}$.  In $\mathbf{k}$-space, this symmetry can be realized by the action of an operator $\hat{Q}$ as follows:
\begin{align}
\hat{Q} \Delta\rho(\mathbf{k},\s) \hat{Q}^{\dag} = -\Delta\rho(-\mathbf{k},\s) {,}
\end{align}
where $\hat{Q}=i\hat{\tau}_2$ in terms of the Pauli matrices $\hat{\tau}_i$ in the sublattice basis.  Hence, the eigenvalues $f_{n,\mathbf{k},\s}$ have particle-hole symmetry with respect to $f_{n,\mathbf{k},\s}=\f{1}{2}$, which implies the conditions $f_{1\s}+f_{6\s}=1$, $f_{2\s}+f_{5\s}=1$ and $f_{3\s}+f_{4\s}=1$ referred to in the main text.
